# MICE: the Muon Ionization Cooling Experiment
# Step I: First Measurement of Emittance with Particle Physics Detectors

U. Bravar
*Space Science Center, University of New Hampshire, Durham, NH 03824, USA*

M. Bogomilov, Y. Karadzhov, D. Kolev, I. Russinov, R. Tsenov
*Dept. of Atomic Physics, University of Sofia, 5 James Bourchier Blvd, BG-1164 Sofia, Bulgaria*

L. Wang, F. Y. Xu, S. X. Zheng
*Inst. for Cryogenic & Superconductivity, Harbin Institute of Technology, Harbin, 150080, PR China*

R. Bertoni, M. Bonesini, R. Mazza
*Sez. INFN Milano Bicocca, Dipartimento di Fisica G. Occhialini, piazza Scienza 3, Milano, Italy*

V. Palladino
*Sez. INFN Napoli e Università Federico II, Napoli, Italy*

G. Cecchet, A. de Bari
*Sez. INFN Pavia e Dipartimento di Fisica Nucleare e Teorica, via Bassi 6, Pavia, Italy*

M. Capponi, A. Iaciofano, D. Orestano, F. Pastore, L. Tortora
*Sez. INFN Roma Tre e Dipartimento di Fisica, via della Vasca Navale 84, Roma, Italy*

S. Ishimoto, S. Suzuki, K. Yoshimura
*KEK, Institute of Particle and Nuclear Studies, Tsukuba, Ibaraki, Japan*

Y. Mori
*Kyoto University Research Reactor Institute, Kumatori-cho Sennan-gun, Osaka 590-0494, Japan*

Y. Kuno, H. Sakamoto, A. Sato, T. Yano, M. Yoshida
*Osaka University, Graduate School of Science, Dept. of Physics, Toyonaka, Osaka, Japan*

F. Filthaut*
*NIKHEF, Amsterdam, The Netherlands*
**also at Radboud University Nijmegen, Nijmegen, The Netherlands*

M. Vretenar, S. Ramberger
*CERN, Geneva, Switzerland*

A. Blondel, F. Cadoux, F. Masciocchi, J.S. Graulich, V. Verguilov, H. Wisting
*DPNC, Section de Physique, Université de Genève, Switzerland*

C. Petitjean
*Paul Scherrer Institut, CH 5232 Villigen PSI, Switzerland*

R. Seviour
*The Cockcroft Inst., Daresbury Science & Innovation Campus, Daresbury, Warrington WA4 4AD, UK*

M. Ellis*, P. Kyberd, M. Littlefield, J.J. Nebrensky
*Brunel University, Uxbridge, Middlesex UB8 3PH, UK*
**now at Westpac Institutional Bank, Sydney, NSW, Australia*

D. Forrest, F. J. P. Soler, K. Walaron*
*School of Physics & Astronomy, Kelvin Bldg., The University of Glasgow, Glasgow, G12 8QQ, UK*
**also at Imperial College, London*

P. Cooke, R. Gamet
*Department of Physics, University of Liverpool, Oxford St, Liverpool L69 7ZE, UK*

A. Alecou, M. Apollonio*, G. Barber, A. Dobbs, P. Dornan, A. Fish, R. Hare, A. Jamdagni, V. Kasey, M. Khaleeq, K. Long, J. Pasternak, H. Sakamoto, T. Sashalmi
*Imperial College of Science, Technology and Medicine, Prince Consort Road, London SW7 2BW, UK*
**now at Diamond Light Source, Harwell Science & Innovation Campus, Didcot, Oxfordshire, UK*

V. Blackmore, J. Cobb, W. Lau, M. Rayner, C.D. Tunnell, H. Witte, S. Yang
*Dept. of Physics, University of Oxford, Denys Wilkinson Bldg., Keble Road, Oxford OX1 3RH, UK*

J. Alexander, G. Charnley, S. Griffiths, B. Martlew, A. Moss, I. Mullacrane, A. Oats, S. York
*STFC Daresbury Laboratory, Daresbury, Warrington, Cheshire, WA4 4AD, UK*

R. Apsimon, R.J. Alexander, P. Barclay, D. E. Baynham, T. W. Bradshaw, M. Courthold, R. Edgecock, T. Hayler, M. Hills, T. Jones, N. McCubbin, W. J. Murray, C. Nelson, A. Nicholls, P. R. Norton, C. Prior, J. H. Rochford, C. Rogers, W. Spensley, K. Tilley
*STFC Rutherford Appleton Laboratory, Chilton, Didcot, Oxfordshire, OX11 0QX, UK*

C.N. Booth, P. Hodgson, R. Nicholson, E. Overton, M. Robinson, P. Smith
*Department of Physics and Astronomy, University of Sheffield, Sheffield S3 7RH, UK*

D. Adey, J. Back, S. Boyd, P. Harrison
*Department of Physics, University of Warwick, Coventry, CV4 7AL, UK*

J. Norem
*Argonne National Laboratory, 9700 S. Cass Avenue, Argonne, IL 60439, USA*

A.D. Bross, S. Geer, A. Moretti, D. Neuffer, M. Popovic, Z. Qian, R. Raja, R. Stefanski
*Fermilab, P.O. Box 500, Batavia, IL 60510-0500, USA*

M.A.C. Cummings, T.J. Roberts
*Muons Inc., Batavia, IL 60510, USA*

A. DeMello, M.A. Green, D. Li, A.M. Sessler, S. Virostek, M.S. Zisman
*Lawrence Berkeley National Laboratory, Berkeley, CA 94720, USA*

B. Freemire, P. Hanlet, D. Huang, G. Kafka, D.M. Kaplan, P. Snopok, Y. Torun
*Illinois Institute of Technology, 3101 S. Dearborn St., Chicago, IL 60616, USA*

Y. Onel
*University of Iowa, Iowa City, IA52242, USA*

D. Cline, K. Lee, Y. Fukui, X. Yang
*UCLA Physics Department, Los Angeles, CA 90024, USA*

R.A. Rimmer
*Jefferson Lab, 12000 Jefferson Avenue, Newport News, VA 23606, USA*

L.M. Cremaldi, T.L. Hart, D.J. Summers
*University of Mississippi, Oxford, MS 38677, USA*

L. Coney, R. Fletcher, G.G. Hanson, C. Heidt
*University of California Riverside, Riverside, CA 92521-0413 USA*

J. Gallardo, S. Kahn, H. Kirk, R.B. Palmer
*Brookhaven National Laboratory, Upton, NY 11973-5000, USA*

The Muon Ionization Cooling Experiment (MICE) is a strategic R&D project intended to demonstrate the only practical solution to providing high brilliance beams necessary for a neutrino factory or muon collider. MICE is under development at the Rutherford Appleton Laboratory (RAL) in the United Kingdom. It comprises a dedicated beamline to generate a range of input muon emittances and momenta, with time-of-flight and Cherenkov detectors to ensure a pure muon beam. The emittance of the incoming beam will be measured in the upstream magnetic spectrometer with a scintillating fiber tracker. A cooling cell will then follow, alternating energy loss in Liquid Hydrogen ($LH_2$) absorbers to RF cavity acceleration. A second spectrometer, identical to the first, and a second muon identification system will measure the outgoing emittance. In the 2010 run at RAL the muon beamline and most detectors were fully commissioned and a first measurement of the emittance of the muon beam with particle physics (time-of-flight) detectors was performed. The analysis of these data was recently completed and is discussed in this paper. Future steps for MICE, where beam emittance and emittance reduction (cooling) are to be measured with greater accuracy, are also presented.

## 1. Introduction

A muon ionization cooling channel is an essential ingredient in the design of next-generation intense muon beam accelerators such as the Neutrino Factory and Muon Collider [1], [2]. In the Neutrino Factory, an intense neutrino beam, ideal to measure the neutrino mass hierarchy and leptonic CP violation, is produced from the decay of muons in a storage ring. Similarly, in the Muon Collider, high-luminosity muon beams are used to investigate multi-TeV lepton − anti-lepton collisions. In all realistic designs for either facility, an intense muon beam is generated by producing pions through proton − target interactions, reaccelerating the muons resulting from pion decay and funneling them into a muon storage ring.

Muons resulting from pion decay have a large and diffuse phase space density (i.e. large emittance ε). In order to fit inside a typical accelerating aperture, even the large aperture FFAGs being adopted in the Neutrino Factory and Muon

Collider designs, the transverse size of the beam must be reduced by orders of magnitude and at the same time the muons must fit longitudinally into an RF bucket for acceleration. In other words, the muon beam requires six dimensional cooling.

While a number of beam cooling methodologies have been developed and successfully implemented in the second half of the past century, they all require relatively long timeframes to achieve a sufficient level of emittance reduction and are thus suitable for stable particles only. The short muon lifetime of 2.2 µs requires the development of a radically new approach. Ionization cooling is presently the only practical solution to this problem. Conceptually, in a muon cooling channel the muon beam passes through low-Z absorbers interleaved with accelerating radio frequency (RF) cavities. The beam loses longitudinal and transverse momentum in the absorbers, and regains only the longitudinal component of the momentum in the RF cavities. As a result, the transverse emittance is reduced and the beam is cooled.

Ionization cooling is at this point in time a theoretical concept, which has not been demonstrated in practice so far. The goal of the Muon Ionization Cooling Experiment (MICE), under development at the Rutherford Appleton Laboratory (RAL) near Didcot, Oxfordshire, England, is to assemble, commission and evaluate a complete cell of the ionization cooling channel designed in the US Feasibility Study-II [3]. The cooling cell, made of three liquid hydrogen ($LH_2$) absorbers interleaved with two sets of 201 MHz RF cavity modules, is designed to produce a 10% reduction in transverse emittance for beams with momenta in the range 140 − 240 MeV/c and incoming emittances of 3−10 π mm rad. This reduction must be measured to 1% of itself, i.e. with an absolute accuracy of $10^{-3}$, since any lack of accuracy will be multiplied up many times in a full cooling channel. In order for MICE to achieve this goal, muon parameters will be measured event-by-event. Each muon will be precisely tracked by a scintillating fiber tracking spectrometer [4], both upstream and downstream of the cooling channel, while particle identification detectors will be used for background suppression. The beam emittance will be then reconstructed from a large sample of particles, providing proof of the feasibility of ionization cooling.

## 2. Ionization cooling

Both the Neutrino Factory and the Muon Collider are affected by the need to fit a diffuse muon beam into the aperture of a storage ring before the muons decay. A collider has an additional constraint that the luminosity of the machine is dependent on how tightly the beam is focused. In other words, the phase space volume occupied by the muon beam must be compacted for a Neutrino Factory or Muon Collider to work. According to Liouville's theorem, this cannot be accomplished by conservative forces, such as e.g. a focusing quadrupole, which may reduce beam size, but only at the expense of increasing momentum divergence.

Muon ionization cooling, first proposed in 1981 [5], is equivalent to a thermodynamic compression doing work on a beam under isothermal conditions. It achieves cooling by introducing a dissipative force, ionization energy loss. The transverse phase space coordinates of the muon beam are reduced by passing the beam through an absorber, where the particles lose energy along their direction of motion, and then restoring the lost energy in the longitudinal component of momentum using RF-cavity acceleration. This process is repeated in multiple (~$10^2$) iterations with the components of the cooling channel immersed in a solenoidal magnetic field used for muon beam focusing. At each step, the muon's momentum is reduced and subsequently increased by ~5% of its value. The end product of the cooling channel is a muon beam with the same energy as the initial beam and significantly reduced emittance.

Transverse emittance reduction is best described by the cooling formula:

$$\frac{d\varepsilon_\perp}{dz} = -\frac{\varepsilon_\perp}{\beta^2 E}\left\langle\frac{dE}{dx}\right\rangle + \frac{\beta_\perp (14 MeV)^2}{2\beta^3 E m_\mu X_0} \quad (1)$$

where $\beta_\perp$ is the betatron function of the cooling channel, measured at the absorber, $X_0$ is the radiation length of the absorber material, $\beta = v/c$ and E, $m_\mu$ are the energy and mass of the muon. This expression includes two distinct components, i.e. the cooling component from ionization energy loss (first term on the right-hand side) and the heating component from multiple scattering (second term on right-hand side). So-called equilibrium emittance is reached when cooling equals heating and $d\varepsilon_\perp/dz = 0$. In order for the equilibrium emittance to be as small as possible, the heating term needs to be minimized. This is achieved by adjusting the channel optics to reduce $\beta_\perp$ as much as practical and selecting an absorber material with $X_0$ as large as practical. Since $X_0$ is loosely proportional to $1/Z$, where Z is the atomic number, the absorber of choice for MICE is liquid hydrogen ($LH_2$), with Z=1.

As the transverse phase space is reduced, the longitudinal phase space grows because of three factors:
- The energy loss fluctuations in the absorber (straggling) cause an increased spread in energy.
- The negative slope of the energy loss curve at lower energies also causes an increased spread in energy.
- The particles themselves have a range of velocities so the bunch lengthens over time.

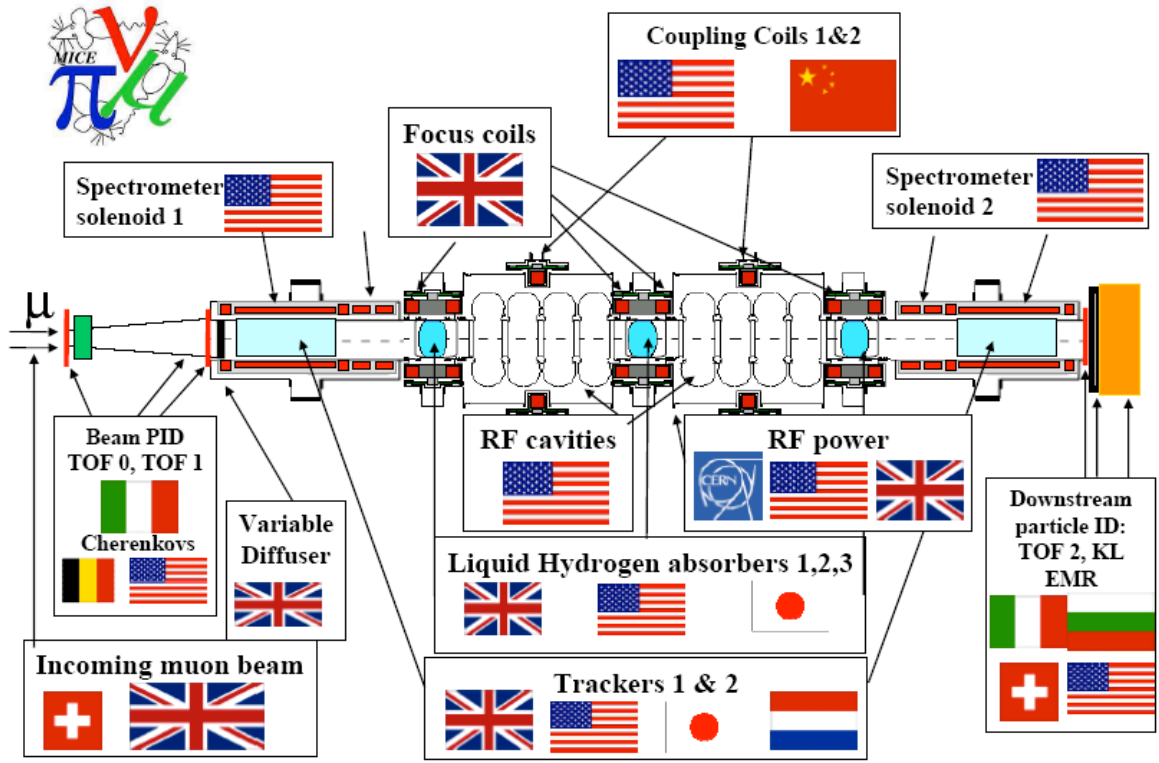

Figure 1: Layout of the MICE experiment, including the cooling cell, spectrometers and particle identification detectors.

To minimize these effects, cooling is carried out near to the energy loss minimum, at a momentum around 200 MeV/c for muons in $LH_2$, since cooling at lower momenta would imply an energy loss curve that rapidly decreases with energy (increasing straggling). Cooling at higher momenta, while providing a slight benefit because of the positive slope, would require an increase in the RF fields to restore the lost momenta.

To achieve six-dimensional cooling, a reduction of longitudinal emittance is required as well. This is achieved through emittance exchange, by literally exchanging some of the beam's initial longitudinal size for an increase in the transverse size, which is more readily reduced. The technique depends on splitting the path of the particles in the beam depending on their energy. A dipole magnet can bend the beam so that the higher energy particles have a larger radius of curvature. The beam is thus correlated in momentum and position, and can be passed into a wedge shaped absorber such that the particles with higher energy cross a thicker piece of absorber and lose more energy. The result is that the energy spread of the beam is reduced but the width and hence transverse size of the beam is increased.

## 3. MICE

The heart of the MICE experiment is a section of a full cooling channel, made up of three liquid hydrogen absorbers and two sets of RF accelerating cavities, with four cavities per set, immersed in a solenoidal magnetic field. In effect this gives two complete cooling half-cells, resulting in one full cell, with the third $LH_2$ absorber being in place for symmetry reasons and to absorb any dark currents from the RF cavities that would produce X-rays affecting the particle trackers. The cooling channel section has spectrometers and particle identification detectors at each end to measure the properties of the muon beam, which is extracted from the RAL pulsed neutron and muon source (ISIS).

The ISIS beamline consists of a 800 MeV proton synchrotron. This uses a Penning $H^-$ ion source and accelerates the ions along a 665 kV channel before injection into the main ring in pulses 200 μs long. When entering the synchrotron an aluminum oxide stripping foil removes the electrons from the ions, leaving protons, which are put into two bunches and accelerated to the full energy. Prior to extraction the bunches become 100 ns long and are separated by 230 ns. To produce the muon beam for MICE, on selected pulses, a small titanium cylinder target, installed in ISIS in August 2009, is dipped into the outer halo of the proton beam just before extraction. Pions produced in this collision emerge through a thin window towards the muon beamline and are transported first through a bending dipole for momentum selection and then to a decay solenoid in which muons are produced from $\pi^{\pm}$ decay. Another dipole is used to select the momentum of the muons emerging from this solenoid, and a set of quadrupoles for transport to the MICE channel. Finally, the beamline also includes two beam profile and luminosity monitoring stations. The main parts of the MICE experiment proper are shown in Figure 1.

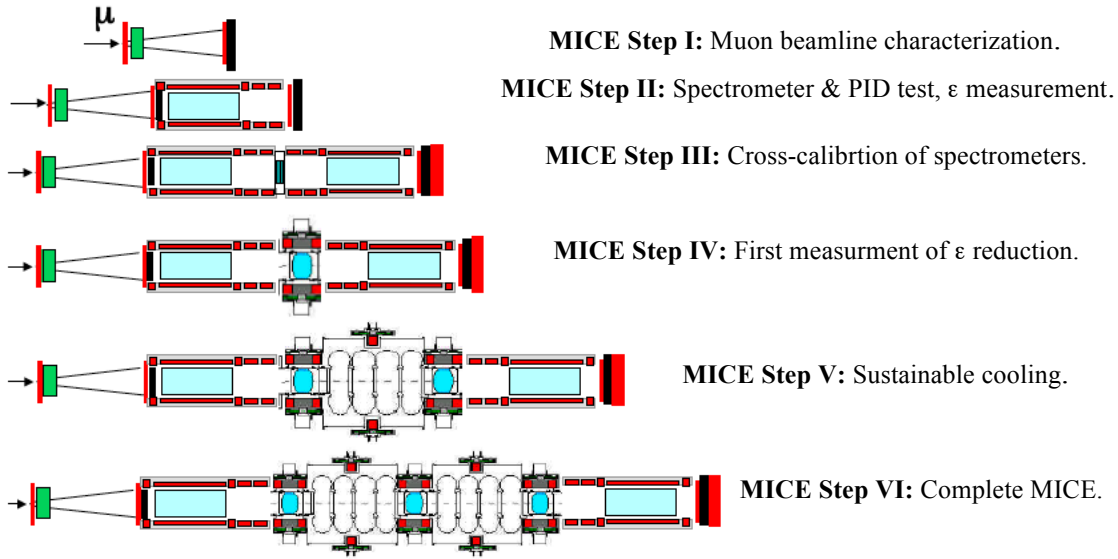

Figure 2: The six steps in the assembly of the MICE experiment.

In the layout of the complete MICE channel, the incoming muons first encounter a time-of-flight station (ToF0), two aerogel Cherenkov detectors and a second time-of-flight station (ToF1), used for particle identification. Next, the muon beam enters the upstream spectrometer, designed to measure the kinematic parameters of each particle. This 1.1 m long component includes an adjustable lead diffuser, to tune the initial emittance of the muon beam, and five scintillating fiber (SciFi) stations, each station consisting of three layers of 350 μm fibers [4], to measure the particle's parameters. This is followed by the first $LH_2$ cell, 35 cm in length and 30 cm in diameter. The $LH_2$ is surrounded by a total of four tapered aluminum windows, two in front and two more in the back, with a varying thickness that decreases to 300 μm at the center of each window. Two RF-cavity modules alternated with two more $LH_2$ cells come next, followed by the downstream spectrometer, identical in construction to the upstream spectrometer save the diffuser. The length of the entire cooling channel section of MICE (i.e. $LH_2$ cells and RF cavities) is ~6 m. The downstream spectrometer is followed by a third time-of-flight station (ToF2) and a calorimeter consisting of two modules, labeled respectively KL (Kloe Light calorimeter) and EMR (Electron–Muon Ranger), used again for particle identification.

Each spectrometer is surrounded by five superconducting coils, three of them providing a uniform solenoidal magnetic field of 4 T for the measurement of muon parameters and the remaining two for beam optics tuning, namely to match the muon beam between the spectrometer and the cooling channel. Eight more superconducting coils with a solenoidal field surround the cooling channel section: two focus coils lay around each $LH_2$ cell and a coupling coil is used half-way through each RF module [6], [7]. The magnetic field intensities can be varied to produce the desired profile of the $β_⊥$-function along the MICE cooling cell and in particular in the $LH_2$ absorbers. We plan to test the cooling cell with $β_⊥$ ranging from 7 cm to 42 cm.

## 4. MICE Progress

The MICE experiment was planned to evolve in six steps, as illustrated in Figure 2, with each step representing a specific milestone. Muon beam data acquisition starts from the very beginning of Step I and cooling channel components are added at various times until the commissioning of the complete experiment in Step VI. In summary:
- Step I includes only particle identification detectors, aimed at muon beamline characterization.
- The upstream spectrometer is added in Step II, for testing and first precise ε measurement by spectrometer.
- Step III includes both spectrometers, for cross-calibration.
- The first $LH_2$ cell is added in Step IV, which should provide the first measurement of ε reduction.
- One more $LH_2$ cell and a RF section are added to Step V, demonstrating sustainable cooling.
- In the end, Step VI is the final assembly stage of the complete MICE experiment.

The MICE beamline was commissioned in 2009 and Step I, with the exception of part of the EMR calorimeter, was completed in 2010. Step I was exposed to the muon beam in the summer of 2010 during the ISIS User Run, which ended with 340,000 target actuations and the acquisition of over $10^7$ triggers. The remaining components of MICE are presently at varying stages of development, most of them nearing delivery. Due to the completion agenda of each component and the need to fit into the schedule of the ISIS beamline, a decision was made recently to proceed directly from Step I to Step IV, with data acquisition taking place again in the third quarter of 2012. In addition to its own goals, Step IV is expected to accomplish the goals of the missing Steps II and III as well. Finally, the second quarter of 2014 is the target date for Step V and Step VI is expected to follow after the long ISIS shutdown planned from August 2014 until February 2015.

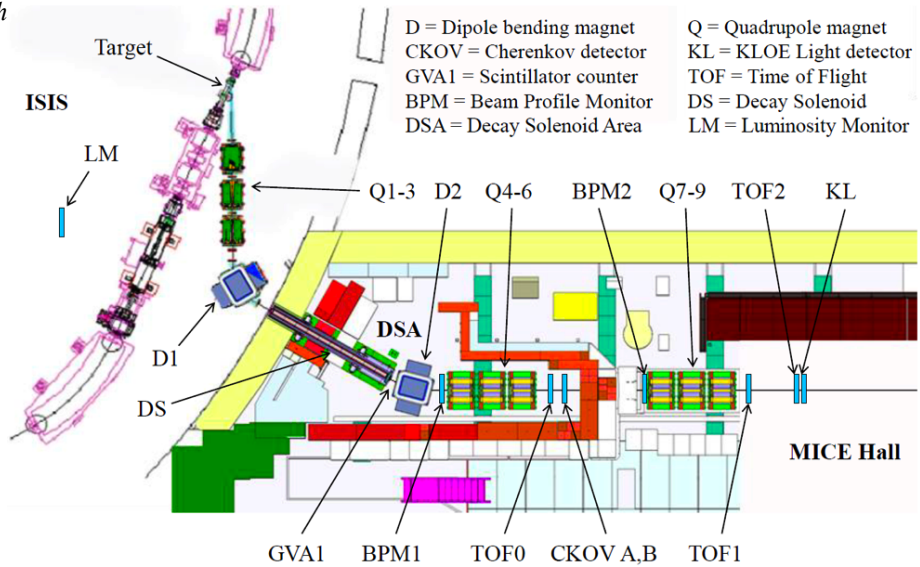

Figure 3: Schematics of MICE Step I as installed in the ISIS muon beamline.

## 5. MICE Step I

The layout of MICE Step I is shown in Figure 3. The goals of this step include the characterization of the ISIS beamline, the calibration of the beamline monitoring detectors and the final commissioning of the particle identification detectors (three ToF stations, two aerogel Cherenkov detectors and one two-module calorimeter), both upstream and downstream of the cooling channel section. There are nine transverse emittance $\varepsilon_\perp$ / longitudinal momentum $p_z$ pairings to be tested by MICE, summarized in Table I. Both the nominal momentum and nominal emittance in this table are taken from the midpoint of the MICE cooling channel, i.e. the center of the $LH_2$ absorber placed in the middle of the experiment. The parameters of the baseline muon beam are $<p_z> = 200$ MeV/c and $\varepsilon_\perp = 6\ \pi$ mm rad.

Table I: Transverse emittance and longitudinal momentum pairs tested by the MICE experiment.

| $<p_z>$=140 MeV/c, $\varepsilon_\perp$=3 π mm rad | $<p_z>$=200 MeV/c, $\varepsilon_\perp$=3 π mm rad | $<p_z>$=240 MeV/c, $\varepsilon_\perp$=3 π mm rad |
|---|---|---|
| $<p_z>$=140 MeV/c, $\varepsilon_\perp$=6 π mm rad | $<p_z>$=200 MeV/c, $\varepsilon_\perp$=6 π mm rad | $<p_z>$=240 MeV/c, $\varepsilon_\perp$=6 π mm rad |
| $<p_z>$=140 MeV/c, $\varepsilon_\perp$=10 π mm rad | $<p_z>$=200 MeV/c, $\varepsilon_\perp$=10 π mm rad | $<p_z>$=240 MeV/c, $\varepsilon_\perp$=10 π mm rad |

Muon, pion and electron beams can all be extracted from the ISIS beamline. All three types of particles were injected into MICE Step I during the data acquisition campaign of 2010. In particular, both $\mu^+$ and $\mu^-$ muon beams were generated for each of the ($<p_z>$, $\varepsilon_\perp$) pairs from Table I.

These data were used to completely characterize the MICE muon beam. Beam composition and purity, as well as data rate and beam quality were determined. In addition, the three ToF stations were used for a preliminary measurement of the emittance of the muon beam. While significantly less accurate than the MICE spectrometers, and unsuitable for achieving the nominal $10^{-3}$ precision required of MICE, these measurements were nevertheless crucial in characterizing the beam parameters.

The ToF detector system, designed primarily to provide time-of-flight measurements to distinguish muons from electrons and pions in the MICE beam, includes three stations (ToF0, ToF1 and ToF2). Each station consists of two orthogonal layers of scintillator bars (see Figure 4). The time resolutions of the three ToFs were determined to equal 51 ps, 58 ps and

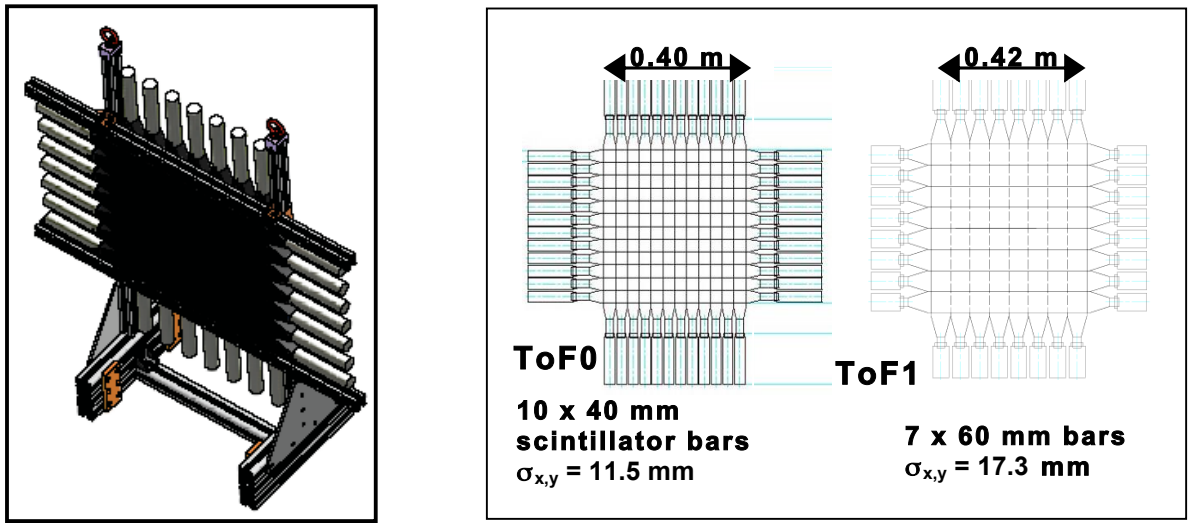

Figure 4: ToF detector design. Pictorial 3-D view (left) and schematic structure (right).

52 ps, respectively [8]. For emittance measurements, a sample of good muons was selected using timing information from ToF0 and ToF1. Under normal circumstances, emittance would be determined using the r.m.s. transverse beam size from three beam profile measurements. However, this is not possible in MICE where there is a large spread in $p_z$, approximately 100 MeV/c for a 240 MeV/c muon beam. Therefore, an alternate approach was developed for the MICE novel single-particle, muon-by-muon, method of emittance measurement, consisting of the following steps:
- Measure x, y, and t at the first two ToF stations (ToF0, ToF1).
- Identify muons using time-of-flight.
- Use momentum-dependent transfer matrices to map the particle motion from ToF0 to ToF1.
- Make an initial estimate of path length and longitudinal momentum.
- Use the computed $p_z$ value and the transfer map to improve the path length estimate, and recalculate $p_z$.
- Iterate the previous two steps until $p_z$ converges.
- Determine trace space emittances (x, $x' = p_x/p_z$) and (y, $y' = p_y/p_z$) at ToF0 and ToF1.

Normalized transverse 2-D beam emittances in both transverse coordinates x and y can then be calculated from the trace space emittances.

## 6. Emittance Measurement in MICE

As stated above, timing information from ToF0 and ToF1 detectors was used in the beginning to select a muon sample and reject background particles such as pions and electrons (Figure 5). ToF0 and ToF1 were also used to determine the transverse coordinates of each particle at each station. The precise timing information from the photo-multiplier (PMT) signals on both ends of each scintillating bar provided a spatial resolution of around 1 cm, better than the segmentation of the bars themselves and better than the resolution achieved from pulse height.

Once the initial and final particle positions were measured, the muon track through the MICE channel was estimated. Each muon had to traverse the two aerogel Cherenkov detectors, several drift spaces, and the last quadrupole triplet in the MICE beamline while crossing the 7.71 m distance between ToF0 and ToF1. A pion contamination of <1%, already predicted by our Geant4-based [9] Monte Carlo simulations [10], was observed independently both by the Cherenkov detectors and by the ToF stations.

Next, products of momentum-dependent transfer matrices were used to reconstruct the muon's track through the drift spaces and quadrupole magnets in the present configuration of the MICE channel. This was an iterative process where the initial path estimate was assumed to be a straight line. The initial value of the muon momentum was then computed from:

$$\frac{p_z}{E} = \frac{s}{\Delta t} \qquad (2)$$

where s, $\Delta t$ are the tracklenght and time-of-flight, respectively. The initial version of the transfer matrix was then determined and the muon tracked through each drift space and magnet using a thick edge quadrupole model. A new value of the tracklength was thus computed and a new value for $p_z$ recalculated. This process was repeated until convergence was achieved, typically after five or fewer iterations. Once the true value of $p_z$ and correct transfer matrix were known, the phase space angles $x' = p_x/p_z$ and $y' = p_y/p_z$ were deduced from the initial and final muon positions.

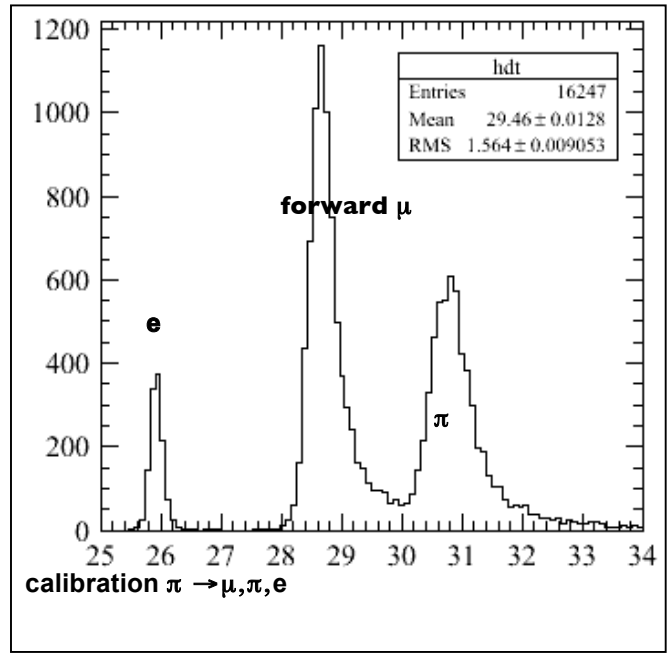

Figure 5: Identification of π, μ, e⁻ particles from the measured time-of-flight between ToF0 and ToF 1.

After a sufficient number of events was collected, the horizontal and vertical trace space, $(x,x')$ and $(y,y')$, distributions were plotted and the transverse beam emittances $\varepsilon_x$ and $\varepsilon_y$ calculated. This emittance measurement technique was applied to all of the MICE muon beam configurations from Table I, for both $\mu^+$ and $\mu^-$, with the exception of the $\varepsilon = 3$ π mm rad, $\langle p_z \rangle$ = 140 MeV/c matrix element, which did not have a sufficiently large muon sample to produce a statistically significant result and needs to be re-run at a later date.

Figure 6 presents the vertical trace space $(y,y')$ scatterplots for the baseline $\mu^-$ beam, with $\varepsilon = 6$ π mm rad and $\langle p_z \rangle$ = 200 MeV/c, where Geant4-based Monte Carlo simulations are compared to experimental data. Similar plots were obtained for the horizontal trace space $(x,x')$. The first plot on the left-hand side displays true Monte Carlo coordinates, obtained by retrieving the information on the particle's actual track from the simulated data. The second plot in the center of the figure presents reconstructed Monte Carlo coordinates, where the $(y,y')$ pixels were obtained from the same Monte Carlo data sample by applying equation (2) and the remaining steps from the previous paragraphs for emittance computation to the detector responses in the simulated data. Minor but non-negligible differences are clearly observable. In particular, our track reconstruction procedure skews the true distribution at the edges of the beam where nonlinear effects are significant. The third scatterplot on the right-hand side of Figure 6, obtained from experimental data, is similar to the one from reconstructed Monte Carlo data in the center of the figure and displays the same skewed shape for high amplitude particles.

## 7. Conclusions

The MICE muon beamline was successfully commissioned and the first measurements of the ISIS muon beam emittance were completed in 2010, using a technique based on ToF measurements uniquely developed for the MICE channel. This milestone represents an important step forward in the development of ionization cooling technology.

Most of the presently missing components of the MICE experiment are in their final developmental stages. A new data sample will be acquired in the third quarter of 2012 with Step IV from the MICE assembly schedule (Figure 2), i.e. with an upgraded apparatus that includes two tracking spectrometers and a cooling LH$_2$ absorber. The trackers, calibrated with cosmic ray data, provide a measured track residual of 661±2 μm, with a 470 μm space point resolution. They will precisely measure the transverse coordinates and longitudinal momentum $p_z$ for each particle, as well as beam emittance into and out of the MICE cooling station. They will also measure the two components of the transverse momentum of each particle to better than 2 MeV/c. The LH$_2$ absorber in Step IV will allow for the first observation of emittance reduction.

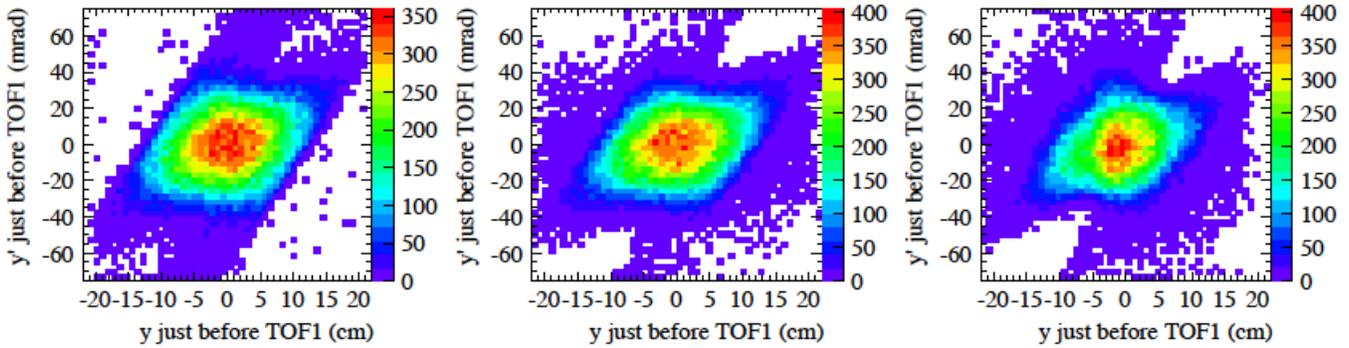

Figure 6: Vertical trace space emittance at ToF1, (y,y') coordinates. True values from Monte Carlo data (left), reconstructed Monte Carlo data (center) and reconstructed experimental data (right) are compared.

MICE will then evolve into a full cooling cell. Step-by-step, two more LH$_2$ absorbers and two RF-cavity sections will be added, with the full experiment targeted to be in operation by 2015. MICE Step VI will demonstrate that the sustainable cooling performance of the Feasibility Study-II cell is achievable in practice and measure emittance redution with the required accuracy of $10^{-3}$. Results from the MICE experiment will then become crucial in the engineering design of the cooling channel of any future Neutrino Factory and/or Muon Collider.

## Acknowledgements

This work was supported by NSF grant PHY-0842798.